\documentclass[showpacs,preprintnumbers,amsmath,amssymb,prl,twocolumn]{revtex4}

\input epsf.tex

\def\comment#1{}

\scriptsize

\def\aut#1{#1}
\begin{document}
\sloppy
\def\bpi{\begin{picture}}
\def\epi{\end{picture}}

\author{H.~Kleinert%\thanks{E-mail kleinert@physik.fu-berlin.de,~\newline
%URL www.physik.fu-berlin.de/\~{}kleinert}
                   \\
         Freie Universit\"at Berlin\\
          Institut f\"ur Theoretische Physik\\
          Arnimallee14, D-14195 Berlin
     }
\title{
Stiff  Quantum Polymers
}
\begin{abstract}
At ultralow temperatures, polymers exhibit quantum behavior,
which is calculated
here
for the  second and
fourth
 moments
 of the end-to-end distribution
 in the large-stiffness regime.
  The result should be measurable
for polymers in
wide optical traps.
\end{abstract}
%%%%%%

\maketitle

{}~\\
{\bf 1.}
Present-day laser techniques
make it
 possible to build
optical traps, and lattices of traps, in which one can
host a variety of atoms or molecules
and study their
behavior at
 very low temperatures.
Gases of bosons, fermions, and
their simple bound states
have been investigated
in this way with interesting insights into
the quantum physics of many-body systems \cite{BOOK}.
In this note we would like to propose
to use these traps
for the study of the
quantum behavior
of stiff polymers.
The low temperatures can be reached
by buffergas cooling with He
which
permits reaching temperatures of the order of mK.
This should in be possible, for example,
with
carbohydrates or polyacetylene.
We shall assume the traps to be much wider
than the length of the polymer
so that
we may ignore the distortions
coming from the trap potential.

The end-to-end distribution
$P_L({\bf R})$
of a polymer
 of length $L$
contains information
on various experimentally observable properties, in particular
 the moments
\begin{eqnarray}    \!\!\!\!\!\!\!\!\!\!\!\!
\left\langle R^m\right\rangle = S_D\int_0^\infty
 dR\,R^{D-1}\,R^m\,P_L({\bf R}),
\end{eqnarray}
where $S_D=2\pi^{D/2}/\Gamma(D/2)$ is the surface of a
unit sphere in $D$ dimensions.
The classical temperature behavior of these moments
is well known \cite{YA,PI}.
Here we
shall calculate
the modifications
caused by
 quantum fluctuations.

Let us briefly recall
the calculation of the
classical
end-to-end distribution in
the Kratky-Porod chain with $N$ links
of length $a$
in $D$ dimensions \cite{YA,PI}. Its
 bending energy is
%innthe continuum limit is given by
\begin{eqnarray} \label{15.81}
E^N_{\rm bend} = \frac{\kappa a }{2} \sum_{n=1}^{N-1} (\nabla {\bf u}_n
         )^2,
\end{eqnarray}
where
$ \kappa $  the stiffness,
 ${\bf u}_n$ are
unit vectors on a
sphere
in
$D$ dimensions
specifying the directions of the polymer links, and $\nabla {\bf u}_n
\equiv
 ({\bf u}_{n+1}
 -{\bf u}_{n})/a
 $ is the
difference between neighboring ${\bf u}_n$'s.
%the so-called lattice derivative.
%lattice derivative from link to link.
%(whose dual is $\bar \nabla\equiv  ({\bf u}_{n}  -{\bf u}_{n-1})/a$).
The initial and final
link directions have
a  distribution
\begin{eqnarray}
P({\bf u}_2,
        {\bf u}_1| L)\equiv
({\bf u}_bL|{\bf u}_a0)=
\int {\cal D}^D{\bf u}  \,
         e^{ - \beta E^N_{\rm bend}} ,
\label{15.82ee}
 \end{eqnarray}
where ${\cal D}^D{\bf u}$ is
the product of integrals over the unit spheres of ${\bf u}_n$ $(n=2,\dots,N-1)$, and
$ \beta \equiv 1/k_BT$ ($T$=temperature, $k_B$= Boltzmann constant).
The normalization
is irrelevant and will be fixed at the end.
%We use natural units  where
%the Boltzmann constant $k_B$ is equal to unity
%so that $ \beta $ is the inverse temperature.

{}~\\
{\bf 2.}
If $L$ denotes the length of the polymer, the bending energy
reads $
E_{\rm bend}^L =
{\kappa } \int_{0}^{L}
 ds \left({\partial_s {\bf u}}\right) ^2/2$.
Then the probability (\ref{15.82ee})
coincides with the Euclidean path integral
of a particle on the surface of a unit sphere.
The end-to-end distance
in space is ${\bf R}=\int_0^L ds\,{\bf u}(s)$, and its distribution
is given by the path integral
\begin{eqnarray}
P_L({\bf R}) \!\!\!
&\propto&\!\!\!
\int {\cal D}^D{\bf u}\,
 \delta^{(D)}
 \left( {\bf R}
-L{\bf u}_0\right)
  e^{ -
{\bar \kappa  }
          \small\int_0^L \!ds\,
  {\bf u}'{}^2(s )/2} ,
\label{15.82n2}
 \end{eqnarray}
where
$\bar  \kappa \equiv  \beta  \kappa $
and
 ${\bf u}_0\equiv L^{-1}\int _0^Lds\,{\bf u}(s)$.
%is the average position of the path in ${\bf u}$-space.
Introducing the dimensionless vectors  ${\bf q}^T$ transverse
to ${\bf R}$, we parametrize ${\bf u}$ as
$({\bf q}, \sqrt{1-{{\bf q}}^2})$ and see
that the $ \delta $-function
enforces
\begin{equation}
\int_0^L ds\, {\bf q}(s)=0,~
R=L-\int_0^L ds[{\bf q}^2(s)/2+\dots].
\label{@REQ}\end{equation}

At large stiffness,
the
 distribution
can be calculated from the one-loop approximation
to the path integral which
leads to
the Fourier integral \cite{PI,FR}
\begin{eqnarray}\!\!\!\!\!
P_{L, \beta }({\bf R}) ~
&\displaystyle\mathop{\propto}_{{\rm small} \,\beta}& ~
\int _{-i\infty }^{i\infty }
\frac{dk^2 }{2\pi i}
 e^{ \beta   \kappa k^2
        \left(L-R
\right)}\,F_{L, 0 }(  k^2L^2 )
\label{15.82nc3}
 \end{eqnarray}
where
$F_{L,0}(k^2L^2)$ is the partition function
\begin{eqnarray}\!\!\!\!\!\!\!\!\!\!\!\!\!\!
F_{L,0}(  k^2L^2  )\!\!\!&\equiv &\!\!\!\!\int_{\rm NBC} \!{\cal D}
\hspace{0pt}'{\hspace{1pt}}^{D-1}{\bf q}^T
     e^{ -( \beta \kappa a/2)\sum_{n=1}^{N}
\left[{{(\nabla{\bf q})^T_n}}{}^2
\!+  k^2      {{\bf q}_n^T}^2\right]
}  \!\!\!\!\!\!\!\!\!\!\!\!\!    \nonumber \\
&&
\!\!  \!\! \!\!\!\!\!\!\!\!
\!\!\!\!\!\!\!\!\!\!\!\propto
\left[  \frac
{ \prod_{n=1}^\infty  |K_n|^2}
{ \prod_{n=1}^\infty ( |K _n|^2+  k^2)}\right] ^{\frac{D-1}{2}}
\!\!\!\!= \left(  \frac{N \sinh \tilde k a }{\sinh \tilde k L}\right)
^{\frac{D-1}{2}}\!,
 \!\!\!\!\!\!
  \!\!\!\!\!\!\!\!\!\!\!\!\!\!\!\!
  \!\!\!\!\!\!\!\!
 \!\!\!\!\!\! \!\!\!\!\!\!\!\!
  \!\!\!\!\!\!\!\!
 \!\!\!\!\!\!  \!\!\!\!\!\!\!\!
 \!\!\!\!\!\!  \!\!\!\!\!\!\!\!
\label{15.82nc2}
 \end{eqnarray}
with $\tilde k$ defined by
$\sinh \tilde ka=ka$ \cite{PIFL}.
 The symbol NBC indicates that the
open ends of the path integral
may be accounted for by Neumann boundary conditions \cite{PI3}.

For a classical polymer,
we may use the model in the continuum limit
where
$a\rightarrow 0$. Then
 ${\bf u} _n$ is replaced by
the tangent vector ${\bf u}(s)=\partial _s{\bf x}(s)$
of the space curve ${\bf x}(s)$ of the polymer,
where $s$ is the distance of the link from one of the endpoints
measured along the polymer.
In this limit,
$\tilde kL$ coincides with $kL\equiv \bar k$,
 and     the right-hand side of (\ref{15.82nc2})
can be expanded as
in a power series
of $\bar k$:
{\begin{eqnarray}
F_{L,0}(\bar k^2)\hspace{-1pt}=
 \hspace{-1pt}
1-\frac{D-1}{ 2^2\cdot 3}\bar k^2+
\frac{(D-1)(5D-1)}{2^5\cdot3^2\cdot 5}\bar k^4
+\dots\hspace{1pt}.
\label{@15.ser}\end{eqnarray}
}%
Inserting this into
(\ref{15.82nc3}) and setting $r\equiv R/L$,
we may calculate  the unnormalized moments
$\langle r^m\rangle=\int dr \,r^{D-1+m} P_{L, \beta }$
 from the integrals
\begin{equation}
\langle r^m\rangle=\int dz \,(1+z)^{D-1+m} f(\hat k^2)  \delta (z),
\label{@}\end{equation}
where $\hat k^2L^2$ is the differential operator
 $-(L/ \beta  \kappa )\partial _z=[-2l/(D-1)]\partial _z$,
and $l\equiv (D-1)L/ \beta  \kappa $ is the {\em flexibility\/} of the polymer.
From this we find
\begin{eqnarray}
&&
\!
\!\!\!\!
\!\!\!\! \!
\left\langle r^0\right\rangle\!=\!{\cal N}\left[ 1\!-\!
\frac{D\!-\!1}{6}l\,\hspace{1pt}
\!+\!\frac{(5D\!-\!1)(D\!-\!2)}{360}l^2
\!+\!\dots\right]\!,
\nonumber \\
&&
\!\!
\!\!\!\!
\!\!\!\!
\left\langle r^2\right\rangle\!=\!{\cal N}\left[1\!-\!\frac{D+1}{6}l
\!+\!\frac{(5D\!-\!1)D(D\!+\!1)}{360(D-1)}l^2
\!+\!\dots\right]\! ,
\label{@UNM} \\
&&
\!\!
\!\!\!\!
\!\!\!\!
\left\langle r^4\right\rangle\!=\!{\cal N}\left[1\!-\!\frac{D+3}{6}\hspace{1pt}l
\!+\!\frac{(5D\!-\!1)(D\!+\!2)(D\!+\!3)}{360(D-1)}l^2
\!+\!\dots\right],
\nonumber \end{eqnarray}
where
${\cal N}$ is some constant.
Dividing these by $ \left\langle r^0\right\rangle$, we
arrive at the normalized moments
\cite{KLM}
\begin{eqnarray}
\!\!
\left\langle r^2\right\rangle\!\!\!&=&\!\!\!
1-\frac{1}{3}l
+\frac{13D\!-\!9}{180(D\!-\!1)}l^2\!
+\dots~ ,~~                     ~~~~~~~~
\label{@MOM2} \\
\!\! \!\!\!\!
\!\!\!\!\!\!\!
\left\langle r^4\right\rangle\!\!\!&=&\!\!\! 1-\frac{2}{3}l
+\,\frac{23D\!-\!11}{90(D\!-\!1)~}l^2\!
+\dots~ .~  \label{@MOM4}
\end{eqnarray}

{}~\\
{\bf 3.}
Quantum effects are now taken
into account by
adding for  each
mass point of the
polymer
at ${\bf x}_n$
a kinetic action
\begin{equation}
{\cal A}_{\rm kin}\equiv
    \frac{M}{2}   \int_0^{\hbar  \beta } dt\,
 [\dot{\bf x}_n(t)]^2,
\label{@ACtIo12A}\end{equation}
where
$M$ is the mass.
Since ${\bf u}_n(t)= \nabla
{\bf x}_n(t)$,
the Euclidean action with
 time $\tau =it$ reads
\begin{eqnarray}
 {\cal A}=
 \frac{\kappa a}{2}\int_0^{\hbar  \beta } d\tau \sum_{n=1}^N
\left[   g^{-2}\left({\partial_\tau \nabla^{-1} {\bf u}}_n\right) ^2+
  \left({\nabla {\bf u}_n}\right) ^2\right] ,
\label{@}\end{eqnarray}
where $g\equiv  \sqrt{ \kappa a/M}$, and
$F_{L,0}( \bar k ^2)$ is replaced by
$F_{L, \beta }( \bar k ^2 )=e^{-(D-1) \Gamma _{ L,\beta }( \bar k ^2 )}$ with
\begin{eqnarray}
 \Gamma _{ L,\beta}(\bar k ^2 )=\frac{1}{2}{\rm Tr}\log[-
g^{-2}\partial  _\tau ^2(\nabla\bar\nabla)^{-1}-\nabla\bar\nabla
+k ^2]      .
\label{@}\end{eqnarray}

The eigenvalues of
$i\partial _\tau $ are
the Matsubara frequencies
$ \omega _m=2\pi m/\hbar  \beta,~(m=0,\pm1,\pm2,\dots\,)$,
leading
to the
finite-temperature generalization of
(\ref{15.82nc2}):
\begin{eqnarray}
F_{L,0}(  \bar k^2  )=
\left[  \frac
{\prod_{m,n}( g^{-2} \omega ^2_m+ |K_n|^4)}
{\prod_{m,n} ( g^{-2} \omega ^2_m+|K _n|^4+  k^2|K _n|^2)}\right] ^{(D-1)/2}\!\!\!\!\!\!\!\!\!\!\!\!\!.
\label{15.82nc2w}
 \end{eqnarray}
% where $m$ runs from $-\infty$ to $\infty$, and $n$ from 1 to $\infty$.
Performing the product over the $m$'s,
we
arrive at
\begin{eqnarray} \!\!\!\!\!
F_{L, \beta }(  k  )
=\prod_{n=1}^\infty
 \left[ \frac{ \sinh  {K_n^2}\,\hbar g \beta /2 }{
\sinh  \sqrt{K_n^4+k^2|K _n|^2}\,\hbar g \beta /2}
\right]^{D-1}\!.
\label{15.82nc22}
 \end{eqnarray}

{}~\\
{\bf 4.}
In the product
(\ref{15.82nc22})
we perform an
expansion
in powers of $\bar k\equiv kL$,
and find
\begin{eqnarray}
F_{L, \beta }( \bar k  )=\exp[(D-1)(f_1\bar k^2+f_2\bar k^4+\dots)]
%=1+(D-1)f_1\bar k^2+(D-1)(f_2+(D-1)f_1^2/2)\bar k^4+dots~
,
\label{@FLT}\end{eqnarray}
where
\begin{eqnarray}
f_1(b)&=&-\frac{b}{4\pi^2} \sum_{n=1}^\infty
\coth \frac{n^2b}{2},\label{@ExPre1}\\
f_2(b)&=&\frac{b^2}{32\pi^4} \sum_{n=1}^\infty
\left[
\frac{2}{bn^2}\coth \frac{n^2b}{2}+\left(\coth^2 \frac{n^2b}{2}-1\right)
 \right] .                                     \nonumber \\
%\frac{1}{n^2}\frac{nb}{2}\coth \frac{nb}{2}.
\label{@ExPre2}
\end{eqnarray}
The parameter $b$
is the
 reduced inverse temperature
$    b\equiv \pi^2\hbar g/k_BTL^2$.

As a cross check of the above results
we go to the high-tem\-pera\-tu\-re limit
where $\coth(n^2b/2)\rightarrow 2/n^2b$
 and thus
$f_1(b)\rightarrow  -1/12,~~
f_2(b)\rightarrow 1/360$.
Inserting these into (\ref{@FLT}),
we recover (\ref{@15.ser}).

Quantum behavior sets in
if $b$ becomes larger than unity.
To estimate when this happens
we measure
the lengths $a,L$ in \AA,
the mass $M$ in units of the proton mass,
the temperature  $T$ in mK,
and
the constant $g$
in units
of \AA$^2$/sec,
we find
that
$b\approx 7.380\times 10^6  \sqrt{ \kappa a/A} /TL^2$,
where $A$ is the atomic number $M$.
In these natural units,
$ \kappa $, $a$, $T$, are of order unity,
experimentalists
should be able
to observe the
quantum behavior
for not too long chains.
% (after reinserting the sound velocity $c$).

%We may now easily evaluate $f_{1,2}(b)$
%numerically  for {\em all\/} $T$.

At very  low temperatures
where quantum effects
become most visible we find the asymptotic behavior
\begin{equation}
f_2(b)\rightarrow
\frac{b}{16\pi^4}\sum_{n=1}^\infty\frac{1}{n^2}
 =\frac{b}{96\pi^2}.
\label{@F20}\end{equation}
In this regime, the sum in $f_1(b)$ diverges linearly.
It is made finite by
remembering that we
are dealing with the
continuum limit of a discrete polymer
with $N=L/a$ links.
Hence we must carry the sum only to $n=N$,
and obtain
\begin{eqnarray}
f_1(b)\rightarrow
 -\frac{b}{4\pi^2}\sum_{n=1}^N{1}
= -\frac{b}{4\pi^2}N.
\label{@F10}\end{eqnarray}
\comment{
As a result we find
\begin{eqnarray} \!\!\!\!\!\!\!
F_{L, \beta }( \bar k ^2 )\rightarrow
1\!-\!\frac{D\!-\!1}{4\pi^2}
b N\bar k^2
\!+\!\frac{(D\!-\!1)^2b^2N^2}{32\pi^4}
  \bar k^4
+\dots,
\label{@}\end{eqnarray}
}
Setting $r-1\equiv z$, we replace
$(\bar k^{2})^n$
in
Eq.~(\ref{@FLT})
%$F_{L, \beta }( \bar k ^2 )$
by
$[-2l/(D-1)]^n\partial^n _z \delta (z)$,
and insert the
resulting
 expansion
into the integral
 $\int dz (1+z)^{D-1+m}$
to
find the unnormalized
moments  of
$ r^0$
,
$ r^2$,
$r^4$
at
 zero temperature. From their ratios
 we obtain
the
normalized moments:
\begin{eqnarray}
\!\!\!\!\!\!\!\!\!\!\!\!\!\langle r^2\rangle\!&=&\!1\!+\!
4lf_1
\!+\!l^2\left(4f_1^2+8\frac{2D-1}{D-1}f_2\right)
 +\dots~,
\label{@r2m0}\\
\!\!\!\!\!\!\!\!\!\!\!\!\!\langle r^4\rangle\!&=&\!1\!+ \!
8lf_1
\!+\!l^2\left(24f_1^2+16\frac{2D+1}{D-1}f_2\right)
 +\dots~.
\label{@r4m0}\end{eqnarray}
From these we find
\begin{eqnarray}
\!\!\!\!\!\!\!\!\!\!\!\!\!\langle 1-r\rangle\!&=&\!-\!
2lf_1
\!-\!8l^2f_2
 +\dots~,
\label{@r2m01}\\
\!\!\!\!\!\!\!\!\!\!\!\!\!\langle (1-r)^2\rangle\!&=&\!
l^2\left(
4f^2_1
+\frac{8}{D-1}f_2\right)
 +\dots~,
\label{@r4m0}\end{eqnarray}
and
the cumulant
\begin{eqnarray}
\langle (1-r)^2\rangle _c=l^2\frac{8}{D-1}f_2-32l^3 f_1f_2-64l^4f_2^2\!+\dots\,.
\label{@SecM}\end{eqnarray}
Hence we find in the zero-temperature limits
\begin{eqnarray}
\!\!\!\!\!\!\!\!\!\!\!\!\!\langle r^2\rangle\!\approx\!1\!-\!
\frac{bl}{\pi^2}{N}
,~~~\langle r^4\rangle\!=\!1\!- \!
2\frac{bl}{\pi^2}N,
\label{@r4m}\end{eqnarray}
where $bl\equiv (D-1)\hbar c/ \kappa $.
For large $c$,
the polymer
 at zero temperature
 may appear considerably
shorter
than expected from the linear
extrapolation of the
high-temperature behavior
to zero temperature.

The quantum effect can be studied
most easily by measuring
for a polymer
of high stiffness $ \kappa $
the peak value
of $1-r$
which behaves
like
\begin{equation}
\langle 1-r\rangle\approx
-2lf_1(b)=-2(D-1)\frac{\hbar c } \kappa \frac{1}{b}
f_1  (b).
\label{@RESu}\end{equation}
One may plot the function
$    C(b)\equiv   6 \kappa /
(D-1){\hbar c }  \langle 1-r\rangle$, for which our result implies
the behavior shown in Fig.~1 for various link  numbers $N$.

  We challenge experimentalists to detect
this
behavior.

\begin{figure}[h]
\begin{picture}(105.64,120.645)
\def\dst{\displaystyle}
\def\fsz{\footnotesize}
\def\ssz{\tiny}
\def\IncludeEpsImg#1#2#3#4{\renewcommand{\epsfsize}[2]{#3##1}{\epsfbox{#4}}}
\put(-30,0){\IncludeEpsImg{105.64mm}{26.43mm}{.5000}{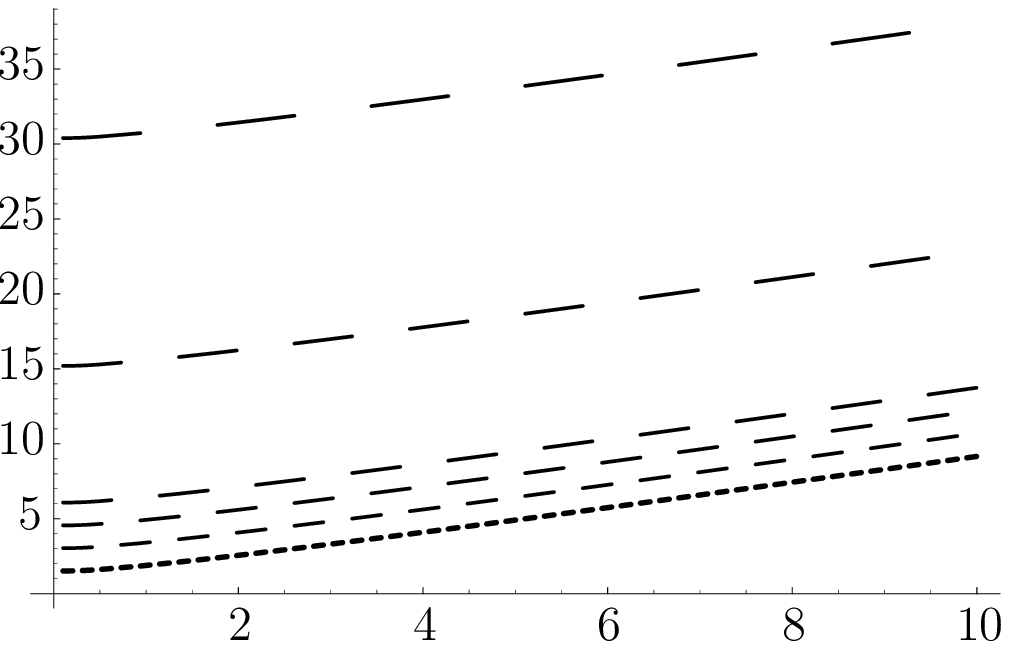}}
%\def\IncludePCXImg#1#2#3#4{\unitlength#3mm\begin{picture}(#1,#2)
%\put(0,#2){\special{em:graph #4}}\end{picture}\vspace{0mm}}
%\IncludePCXImg{100}{100}{1}{pic.pcx}
%\put(0,0){\input c:/emtex/texdraw/grating2}
%\put(0,0){\input c:/emtex/texdraw/gratingf}
\put(125,3.9){\fsz $1/b$}
\put(5,91){\fsz $C(b)$}
%\put(65,30){\fsz classical behavior}
%\put(65.2,35){\vector(-1,1){10}}
\put(115,87){\ssz $N=\phantom{}\,100$}
\put(115,55){\ssz $N=\phantom{0}\,50$}
\put(115,37){\ssz $N=\phantom{0}\,20$}
\put(115,33){\ssz $N=\phantom{0}\,15$}
\put(115,29){\ssz $N=\phantom{0}\,10$}
\put(115,25){\ssz $N=\phantom{00}\,5$}
\end{picture}
\caption[]{Temperature behavior of
$ C(b)\equiv         6 \kappa /
(D-1){\hbar c }  \langle 1-r\rangle$
for various link  numbers $N$. The classical limit of these curves
are their straight-line asymptotes
starting out at the origin
with slope  $(6/\pi^2)\sum_{n=1}^N n^{-2}$.
}
\label{@}\end{figure}
\comment{
where $l$ and $1/b$ are both proportional to $T$.
temperature, we may consider it
as a reduced temperature, and set
 $\hat c_2\equiv lc_2=2[{\hbar c(D-1)}/{4\pi \kappa }]e^{-1/b}$,
 $\hat c_4\equiv lc_4=2[{\hbar c(D-1)}/{4\pi \kappa }]^2e^{-1/b}$.
We challenge experimental physicists
to observe the leading temperature behavior
of $\langle r^2\rangle\approx1-l /3-2e^{-1/b}(\hbar c/  \pi \kappa )$.
}

{}~\\
{\bf 5.}
Further  quantum effects can be observed if the links of the
polymer contain a spin
$S=1/2,\,1,\,3/2,\,2,\dots$ the
link direction.
This can be taken into account by
adding  the kinetic  action
(\ref{@ACtIo12A})
a {\em Berry phase\/}.
For each link ${\bf u}_n(\tau )$, it corresponds to the interaction
of the particle on the surface of a unit sphere in ${\bf u}$ space with a
magnetic monopole
of quantized charge $q$
lying at the center of the sphere \cite{PIM}:
\begin{equation}
{\cal A}_0=
{ \hbar }{S}\sum_{n=1}^{N-1}   \int_0^{\hbar  \beta } d\tau \,
\frac{{\bf n}\times {\bf u}_n(\tau )}{1-{\bf n}\cdot {\bf u}_n(\tau )}\cdot \dot{\bf u}_n(\tau ).
\label{@ACtIo12A}\end{equation}
The irrelevant Dirac string
is chosen to export the magnetic flux
of strength $S$ along the ${\bf n}$ direction to infinity.
This action creates
a radial magnetic field
${\bf B}=-S{\bf u}_n$ on the surface of the sphere.
If we assume  ${\bf R}$ to run along the positive
$z$ direction, the
small
transverse  fluctuations
${\bf q}^T
$ in (\ref{15.82nc2}) will take place near the north pole of the sphere
and receive a
an additional
magnetic
interaction
$\hbar S\sum_{n=1}^{N-1} \int_0^{\hbar  \beta } d\tau \,{\bf q}_n^T\times \dot{\bf q}_n^T/2a$.
This will change each factor in the
product
(\ref{15.82nc22})
to a product of two square roots \cite{PIB}
\begin{eqnarray} \!\!\!\!\!
 \left[ \frac{ \sinh  K_nK^+_n(0)\,\hbar c \beta /2 }
{\sinh   K_nK^+_n(k)\,\hbar c \beta /2}
\right]^{\frac{1}{2}} \!
 \left[ \frac{ \sinh  K_nK^-_n(0)\,\hbar c \beta /2 }
{\sinh   K_nK^-_n(k)\,\hbar c \beta /2}
\right]^{\frac{1}{2}} \!
\label{15.82nc22B}
 \end{eqnarray}
where
\begin{eqnarray}
\!  K^\pm_n(k)\equiv
  \sqrt{K_n^2+k^2+k_S^2}\pm k_S ,~~~~~k_S\equiv \frac{\hbar c S}
{2 \kappa a}.
\label{@}\end{eqnarray}
%
%and $SN$ is the total spin angular momentum of the
the stretched poylmer.
For arbitray
temperatures, this changes
Eqs.~(\ref{@ExPre1})
and
(\ref{@ExPre2})
to
\begin{eqnarray}\!\! \!\!\!\!\!\!\!\!\!\!
f_1(b)&=&-\frac{b}{8\pi^2} \sum_{n=1}^\infty\frac{n}{n_S}
\left(
\coth \frac{n n_S^+ b}{2}
+\coth \frac{n n_S^-b}{2} \right)
,\label{@ExPre}\\  \!\!\!\!\!\! \!\!\!\!\!\!\!\!\!\!
f_2(b)&=&\frac{b^2}{64\pi^4}
\sum_{n=1}^\infty
\left[
\frac{2n}{bn_S^3}
\left(
\coth \frac{nn_S^+ b}{2}
+\coth \frac{nn_S^-b}{2}
\right)\right.
\nonumber \\
&&~\left. +\frac{n^2}{n_S^2}\left(
\coth^2 \frac{nn_S^+ b}{2}
+\coth ^2\frac{nn_S^-b}{2}-2
 \right) \right]     ,
\end{eqnarray}
where
$ n_S^\pm=n_S\pm \kappa _S $,
$ n_S\equiv  \sqrt{n^2+ \kappa _S^2}$,
and $ \kappa _S\equiv   \hbar c SL/2 \pi\kappa  a=k_SL/\pi $.
At high temperatures,
%we obtain the extensions of
%Eqs.~(\ref{@F20}) and (\ref{@F10}) to $ \kappa _S\neq 0$:
these
become
\begin{eqnarray}\!\!\!\!\!\!\!\!\!\!
f^S_1(b)&\rightarrow&
 -\frac{1}{4\pi^2}
\sum_{n=1}^\infty\frac{1}{n_S}\left(
\frac{1}{ n^-_S}
+\frac{1}{ n^+_S}\right)=-\frac{1}{12},
%\nonumber \\&=& -\frac{1}{2\pi^2}
%\sum_{n=1}^\infty\frac{1}{n^2}=-\frac{1}{12}
\label{@F20S}\\\!\!\!\!\!\!\!\!\!\!
f^S_2(b)&\rightarrow&~
\frac{1}{16\pi^4}\sum_{n=1}^\infty\left[
\frac{1}{n_S^3}
\left(
\frac{1}{n^-_S}+
\frac{1}{n^+_S}\right)\right.\nonumber
\\
&&~~~~~\left.+\frac{1}{n_S^2}
\left(
\frac{1}{(n^-_S)^2}+
\frac{1}{(n^+_S)^2}\right)
 \right]=\frac{1}{360}.
%\nonumber \\&=&
%\frac{1}{8\pi^4}\sum_{n=1}^\infty\left[
%\frac{1}{n_S^2n^2} +\frac{1}{n_S^2}
%\frac{n^2+2 \kappa _S^2}{n^2}
\label{@F20S}
\end{eqnarray}
The classical limit is independent
of $ \kappa _S$, as could have been anticipated.

At low temperatures,
we obtain for small $ \kappa _S$ to lowest order
\begin{eqnarray}\!\!\!\!\!\!\!\!\!\!
f^S_1(b)&\rightarrow&
 -\frac{b}{4\pi^2}
\sum_{n=1}^N\frac{n}{n_S}=-\frac{b}{4\pi^2}\left(N-\frac{\pi^2 \kappa _S^2}{12}\right)
%+{\cal O}( \kappa _S^4)
,
\label{@F20SL}\\\!\!\!\!\!\!\!\!\!\!
f^S_2(b)&\rightarrow&~
\frac{b}{16\pi^4}\sum_{n=1}^\infty
\frac{n}{n_S^3}
=\frac{b}{96\pi^2}\left(1-\frac{\pi^2 \kappa _S^2}{10}\right).
\label{@F20SL}
\end{eqnarray}
 Thus $f_1(b)$ depends only very weakly on $ \kappa _S$
so that the curves in
Fig.~1
are practically unchanged  by an extra spin $S$ along the links.
The spin dependence becomes visible
only in measurements
of $f_2(b)$
which can be extracted
from suitable combinations of the moments
$\langle 1-r\rangle$
and $\langle (1-r)^2\rangle_c$
obtained by solving
Eqs.~(\ref{@r2m01}) and
(\ref{@SecM}).

{}~\\
{\bf 7.}
Our discussion has shown
that  at low temperatures
quantum fluctuations
cause  observable
effects
  in  polymers.
We have calculated these
effects   for the lowest moments
$\langle r^2\rangle$ and
$\langle r^4\rangle$
of the end-to-end distribution
for  ordinary
polymers as well as
for polymers in which each link carries a spin $S$.
In the latter case the
polymers are flexible
one-dimensional quantum Heisenberg ferromagnets.
With the
presently
 available traps
and cooling techniques,
experimentalists should be able to detect these effects.

\comment{
~\\
\begin{figure}[h]
\begin{picture}(105.64,120.645)
\def\dst{\displaystyle}
\def\fsz{\footnotesize}
\def\ssz{\tiny}
\def\IncludeEpsImg#1#2#3#4{\renewcommand{\epsfsize}[2]{#3##1}{\epsfbox{#4}}}
\put(-30,0){\IncludeEpsImg{105.64mm}{26.43mm}{.5000}{PLOTL.EPS}}
%\def\IncludePCXImg#1#2#3#4{\unitlength#3mm\begin{picture}(#1,#2)
%\put(0,#2){\special{em:graph #4}}\end{picture}\vspace{0mm}}
%\IncludePCXImg{100}{100}{1}{pic.pcx}
%\put(0,0){\input c:/emtex/texdraw/grating2}
%\put(0,0){\input c:/emtex/texdraw/gratingf}
\put(50,33){\fsz $ \Delta $L${}_S( \kappa _S)$}
%\put(120,79){\fsz $ \kappa _S$}
\put(20,87){\fsz $\kappa _S\equiv   \hbar c SL/2 \pi\kappa  a$}
\end{picture}
\caption[]{Change of the
 parameter L in Eqs.~(\ref{@r2m}) and (\ref{@r4m})
for increasing spin $S$
on each link.
%The abcissa specifies $S$ via the dimensionless
%variable $ kappa _S\equiv   \hbar c SL/2 \pi\kappa  a$.
}
\label{@}\end{figure}
\comment{
where $l$ and $1/b$ are both proportional to $T$.
temperature, we may consider it
as a reduced temperature, and set
 $\hat c_2\equiv lc_2=2[{\hbar c(D-1)}/{4\pi \kappa }]e^{-1/b}$,
 $\hat c_4\equiv lc_4=2[{\hbar c(D-1)}/{4\pi \kappa }]^2e^{-1/b}$.
We challenge experimental physicists
to observe the leading temperature behavior
of $\langle r^2\rangle\approx1-l /3-2e^{-1/b}(\hbar c/  \pi \kappa )$.
}
}

{{}~\\
Acknowledgment: \\
The author is
grateful to I. Bloch,
W. Ketterle, G. Meijer,
F. Nogueira, and T. Pfau
for
 valuable comments.
}

\end{document}